\documentclass[%
twocolumn,
superscriptaddress,
amsmath,amssymb,
prl,
floatfix,
]{revtex4-1}

\usepackage{graphicx}
\usepackage{dcolumn}
\usepackage{bm}
\usepackage{hyperref}
\usepackage{epstopdf}
\usepackage{color}
\usepackage{verbatim}
\graphicspath{{./fig/}}
\usepackage{overpic}

\newcommand{\SL}{\mathrm{SL}(2,\mathbb{R})}
\newcommand{\trm}{\mathrm{Tr}(\mathsf{M})}
\newcommand{\mm}{\mathsf{M}}
\newcommand{\GERRITSCALE}{2.9}

\newcommand{\bi}{\begin{enumerate}}

\newcommand{\ei}{\end{enumerate}}

\newcommand{\ds}{\displaystyle}

 \definecolor{red}{rgb}{1,0,0}
 \definecolor{gre}{rgb}{0,1,0}
 \definecolor{blu}{rgb}{0,0,1}

\begin{document}
\title{Bifurcations of the magnetic axis and the alternating-hyperbolic sawtooth}
\author{C. B. Smiet}
\affiliation{Princeton Plasma Physics Laboratory, Princeton University, Princeton, New Jersey, USA}
\affiliation{Huygens-Kamerlingh Onnes Laboratory, Leiden University, P.O.\ Box 9504, 2300 RA Leiden, The Netherlands}

\author{G. J. Kramer}
\affiliation{Princeton Plasma Physics Laboratory, Princeton University, Princeton, New Jersey, USA}
\author{S. R. Hudson}
\affiliation{Princeton Plasma Physics Laboratory, Princeton University, Princeton, New Jersey, USA}

\begin{abstract}
We present a sawtooth model that explains observations where the central safety factor, $q_0$, stays well below one, which is irreconcilable with current models that predict a reset to $q_0=1$ after the crash.
We identify the structure of the field around the magnetic axis with elements of the Lie group $\SL$ and find a transition to an alternating-hyperbolic geometry when $q_0=2/3$.
This transition is driven by an ideal MHD instability and leads to a chaotic magnetic field near the axis. 
\end{abstract}

\maketitle

The sawtooth oscillation in tokamaks consists of a slow rise in core temperature lasting several to hundreds of ms, followed by a rapid crash lasting $50-200$ $\mu\mathrm{s}$.
It was first observed in 1974~\cite{von1974studies} and has been observed in almost every tokamak since~\cite{chapman2010controlling}.
Understanding and mitigating this temperature-limiting instability is crucial for the success of future tokamak reactors such as ITER.
Existing sawtooth models predict that the crash occurs when the central safety factor $q_0\sim1$ (defined below)
and that the $q=1$ surface is removed.
However, a significantly lower value of $q_0$ is sometimes observed, and some experiments indicate the $q=1$ surface is not removed.
In this paper we present a sawtooth model that predicts a crash to occur by a change in topology of the magnetic field in the core when $q_0=2/3$, and which is consistent with these outlier observations.

The field lines in an axisymmetric tokamak lie on a foliation of nested flux surfaces that confine the plasma.
The single field line at the center of this foliation is called the magnetic axis.
The winding of field lines is quantified by the safety factor, $q$, (inverse of the rotational transform, $\imath$), which quantifies the ratio of toroidal to poloidal winding of a field line on a surface.

Sawtooth models predict a fast-growing ideal instability to occur at some value of $q_0$ ($q$ at the magnetic axis) that rapidly mixes the plasma in the core, and $q_0$ subsequently increases.
Due to the temperature-dependent (Spitzer) resistivity, current re-accumulates near the axis on a (slow) resistive timescale, which decreases $q_0$ until the crash is triggered again.
This leads to a characteristic sawtooth-pattern in diagnostics sensitive to temperature variations that measure near the magnetic axis.

The current leading models, which are the Kadomtsev and the Wesson model, both predict the crash to occur close to $q_0=1$ due to a fast growing mode with 1/1 mode numbers.
In the Kadomtsev model~\cite{kadomtsev1975disruptive}, the crash is triggered when $q_0$ is slightly below $1$ and a $q=1$ surface exists in the plasma.
This configuration is unstable to either an internal kink mode, a resistive tearing instability of the $q=1$ surface, or both~\cite{coppi1976resistive}.
The hot plasma within the $q=1$ surface mixes with cold plasma outside and is deposited in a growing $1/1$ island until all the flux is reconnected.
This model predicts the complete removal of the $q=1$ surface and $q_0=1$ after the crash.
The $q$-profile is flat up to the inversion radius, which lies outside of the pre-crash $q=1$ surface~\cite{kadomtsev1975disruptive}.

The Wesson model~\cite{wesson1986sawtooth} states that the crash is triggered by a quasi-interchange mode that occurs when $q\sim1$ in a region near the axis, also setting $q_0 = 1$ after the crash.
Recent numerical simulations have shown other phenomena in the $q_0\sim1$ regime, where a nonlinearly saturated interchange mode produces a self-organized hybrid stationary state that keeps $q_0$ just above 1~\cite{jardin2015self}.
This has led to the formulation of a new explanation of the sawtooth~\cite{jardin2020new}, in which the saturated 1/1 poloidal flow maintains a flat safety factor profile with $q\gtrsim 1$ until the discharge crosses the stability threshold of the 2/2 mode which triggers a crash through stochastization of the core region.

Though crashes at $q_0\sim1$  are often observed~\cite{weller1987persistent, wroblewski1991determination, nam2018validation}, the above models cannot explain a significant subset of observations which show crashes that occur significantly below $1$, where $q_0$ stays below 1, and where the $q=1$ surface is not removed.
On TEXT Lithium fluorescence imaging was used~\cite{west1987measurement} to measure $q_0=0.7\pm0.1$, and on Tokapole II the differential flux was directly measured through probe insertion~\cite{osborne1982discharges} giving $q_0=0.6-0.8$.
MTX also reported $q_0=0.7-0.8\pm 0.1$ using Faraday rotation imaging~\cite{rice1994poloidal}.
On TEXTOR the measurement was performed using Faraday rotation imaging averaged over many cycles~\cite{soltwisch1986current, soltwisch1988measurement, soltwisch1995sawtooth} resulting in a value of $0.7\pm0.1$, whilst on on TFTR motional stark effect was used~\cite{levinton1993q, yamada1994investigation} to also measure $q_0=0.7\pm 0.1$.
Both of these techniques used simultaneously on JET gave the same result of $0.7\pm 0.1$ \cite{wolf1993comparison}.
Additionally, both MTX\cite{rice1992fifteen} and TEXTOR~\cite{soltwisch1995sawtooth} measured the inversion radius to lie inside the $q=1$ surface which is in contradiction to Kadomtsev.
A clear pattern emerges (for a review see~\cite{park2019newly}): the sawtooth crashes that occur when the safety factor is below one all cluster around the value 0.7.

On Jet, long-lived $(m,n)=(1,1)$ density perturbations, 'snakes', were observed where a deuterium pellet crosses the $q=m/n=1/1$ surface~\cite{weller1987persistent}.
In this operation regime `double snakes' could also be observed, supporting their interpretation as localized density perturbations on the $q=1$ surface.
These snakes can survive sawtooth crashes~\cite{weller1987persistent,gill1992snake}, indicating that the $q=1$ surface is not removed during the crash.
A model for their persistence has been formulated postulating a split-up and re-formation of the snake during post-crash secondary reconnection~\cite{biskamp1994dynamics}, but a simpler explanation is that the $q=1$ surface is simply not removed.
Additionally, Mirnov coil measurements of ellipticity induced Alfv\'en eigenmodes (EAEs) on the JT60U tokamak show modes localized on the $q=1$ surface that are present both before and directly after the crash~\cite{kramer2001magnetic}, also indicating a process that leaves the $q=1$ surface intact.
Many of the crashes analyzed in Ref.~\cite{kramer2001magnetic} show EAE activity consistent with removal of the $q=1$ surface, but some clearly do not.

The above issues are often referred to as the problem of `incomplete reconnection', suggesting that the Kadomtsev process does not run to completion.
Diamagnetic effects can, in theory, stabilize reconnection when the diamagnetic drift speed exceeds a threshold~\cite{beidler2011model}, and such a process would result in a poloidally asymmetric post-crash profile, as can sometimes be observed~\cite{furno2001understanding, pietrzyk1999behaviour}.
A different model that predicts $q_0<1$ after the crash was presented by Kolesnichenko~\cite{kolesnichenko1992sawtooth}, which consists of two reconnection processes, a first in which the field completely reconnects (induced by either the Kadomtsev or Wesson process), and a second which then (partly) reverses the first.
Biskamp and Drake~\cite{biskamp1994dynamics} demonstrated that this reversal can be driven by the energy stored in electron inertia excited by the crash, which after the crash converts to a toroidal current that drives the central safety factor down to a value below 1 but above the pre-crash value.
This model can explain how the safety factor can remain below one after a 1/1 crash, but gives no explanation for why the $q<1$ crash is consistently triggered at $q_0=0.7$.

Different physical processes can lead to similar experimental signatures, namely a rapid crash and equilibration of the central temperature.
We posit that sawtooth crashes that are {\em not} consistent with the Kadomtsev and Wesson models, namely when $q_0$ is not set equal to $1$, are caused by an entirely different mechanism.

We use mathematical group theory to identify a hitherto overlooked bifurcation of the magnetic axis, use ideal magnetohydrodynamcs (MHD) stability calculations to identify the associated unstable mode, and demonstrate that this mode produces magnetic stochasticity.
From this emerges a new model of the sawtooth crash that does not remove the $q=1$ surface, that predicts the crash to occur at $q_0=2/3$, and that explains the observations that are irreconcilable with the models of Kadomtsev and Wesson.

We consider a typical tokamak-like magnetic field in cylindrical coordinates, $(R,\phi,Z)$.
We take the toroidal component to be always positive, so that ${\bm B}\cdot\nabla \phi > 0$, and we assume that there is at least one closed flux surface (a surface where $\bm{B}\cdot \bm{\hat{n}}=0$ with $\bm{\hat{n}}$ the surface normal) somewhere in the domain of interest.
We assume that the magnetic field is continuous, differentiable, and changes continuously in time.

The field line mapping, also called the Poincar\'e mapping, is constructed by integrating once around the torus along the magnetic field.
Integration is started from a point on a prescribed surface, called the Poincar\'e surface, that is transverse to the magnetic field. 
We choose the Poincar\'e section to be the plane $\phi=0$, and write the mapping as ${\bm f}(R_0,Z_0)=(R_1,Z_1)$.
The region of the Poincar\'e section that is bounded by a closed flux surface (a region that is topologically a disk), is mapped to itself under the field line mapping.
The field line mapping is continuous and differentiable.
Because the magnetic field in a tokamak constitutes a Hamiltonian dynamical system~\cite{boozer1985magnetic}, the field line map is the Poincar\'e map or first return map of this dynamical system.

Brouwer's fixed point theorem states that any continuous map from a disk to itself has at least one fixed point~\cite{brouwer1911abbildung}. 
The magnetic axis is an example of a fixed point, as are the points on an intact $q=1$ surface.

At a fixed point we construct the Jacobian matrix of partial derivatives,
\begin{equation}\label{eq:linmap}
   \mathsf{M}=
   \begin{pmatrix}
       \ds \frac{\partial R_1}{\partial R_0}, & \ds \frac{\partial R_1}{\partial Z_0} \\
       \ds \frac{\partial Z_1}{\partial R_0}, & \ds \frac{\partial Z_1}{\partial R_0}
   \end{pmatrix}.
\end{equation}
The matrix $\mm$ describes to first order the behavior of nearby field lines~\cite{greene1968two},
   \begin{equation}\label{eq:eigenvec}
     \bm{f}(\bm{x}_0+\delta {\bm x})\approx \bm{f}({\bm x}_0) + \mm \cdot \delta \bm{x},
    \end{equation}
where $\bm{x}=(R,Z)^T$. 
This matrix can be used to accelerate convergence of iterative methods for finding fixed points and to compute the Lyapunov exponent of the magnetic field lines. 

At a fixed point the divergence free condition on the magnetic field, $\nabla \cdot \bm{B} = 0$, guarantees that this map is area preserving, so $\det(\mathsf{M})=1$.
Therefore $\mathsf{M}\in \SL$, the group of $2\times2$ real-valued matrices with unit determinant.
$\SL$ is a connected simple Lie group, whose elements can be classified into elliptic, hyperbolic and parabolic subsets by their action on the Euclidean plane.

\begin{figure}
   \begin{center}
       \includegraphics[width=8cm]{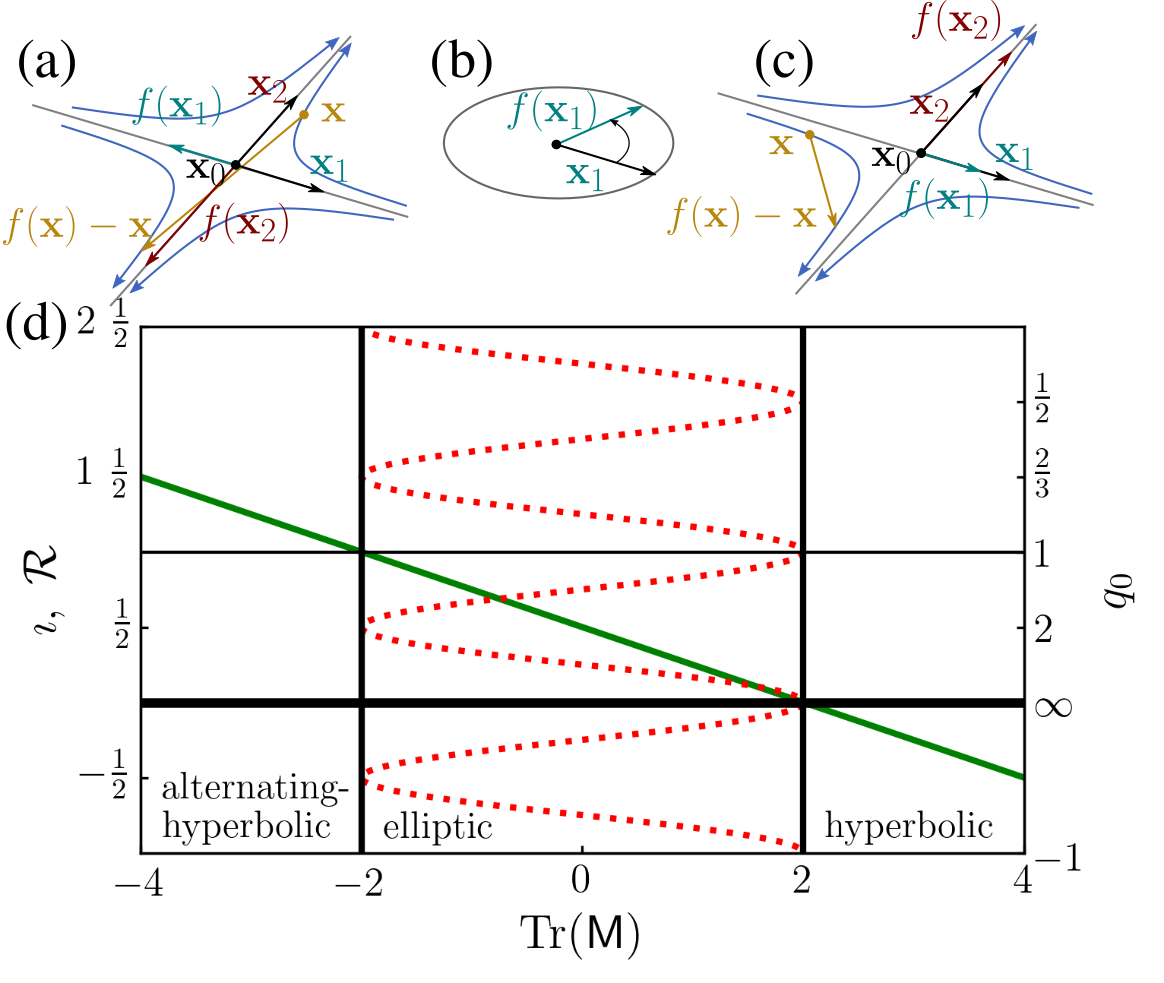}
   \end{center}
   \caption{Classification of the elements of $\SL$.
     (a), (b) and (c): illustration of a typical alternating-hyperbolic, elliptic and hyperbolic element element respectively. Eigenvectors of $\mm$ are shown in black, an example surface invariant under the mapping in blue, and the mapping of select points are illustrated by the colored vectors.\
     (d): Properties of elements of $\SL$ as a function of $\trm$. For the elliptic fixed points the rotational transform $\imath$ (left axes) and safety factor $q_0$ (right axes) is shown in dashed red. Greene's residue (left axes) is shown in green.
   }
\label{fig:sl2r}
\end{figure}

The eigenvalues of $\mathsf{M}$ are solutions to the characteristic polynomial $\lambda_\pm=(\trm \pm \sqrt{\trm^2 - 4})/2$.
Eigenvalues are complex when $|\trm|< 2$ and real when $|\trm|\geq 2$.
When $\trm>2$ the configuration of the field is that of a regular X-point, as illustrated in Fig.~\ref{fig:sl2r}(c), and $\mm$ is a hyperbolic element of $\SL$.
The fixed point is surrounded by hyperbola that are left invariant under this mapping and the (real) eigenvectors of $\mm$ determine the directions in which points, upon consecutive iterations of the field line map, converge to (forming the slow manifold) respectively diverge away from (forming the fast manifold) the fixed point.
When $|\trm|<2$ the configuration of the field is that of an O-point, as illustrated in Fig.~\ref{fig:sl2r}(b), and $\mm$ is an elliptic element of $\SL$.
The fixed point is surrounded by invariant ellipses, which constitute the magnetic surfaces surrounding the fixed point.
When $\trm<-2$ the configuration resembles that of an X-point, but both eigenvalues of $\mm$ are negative.
This mapping sends points on invariant hyperbolic surfaces to the branch on the opposite side of the fixed point, as illustrated in Fig.~\ref{fig:sl2r}(a).
$\mm$ is also a hyperbolic element of $\SL$, but because its map includes a reflection, we call the fixed point an \emph{alternating-hyperbolic} fixed point.
When $|\trm|=2$ the configurations include the identity mapping and shear mappings such as occur at an intact $q=1$ surface.
In this case $\mm$ is called a parabolic element of $\SL$.

The field near an O-point is mapped from an elliptic (magnetic) surface to itself with a certain average rotation.
This determines the local safety factor $q_0$ (up to the multiplicity of the solution) of this fixed point through
   \begin{equation}\label{eq:rottrans}
    \cos(2\pi/q_0) = \tfrac{1}{2}\trm.
   \end{equation}
This relation was given by Greene as $\sin^2(2\pi/2q_0)=\mathcal{R}$~\cite{greene1979method, greene1968two}, where Greene's residue $\mathcal{R}=\frac{1}{2} - \frac{1}{4}\trm$.
There must always be one more fixed point with positive residue than there are fixed points with negative residue\cite{greene1968two, smiet2019mapping}.
The structure of $\SL$, as well as the residue and average angle of rotation are shown in Fig.~\ref{fig:sl2r}(d).

Different configurations of the magnetic field correspond to different elements of this group, and the structure of this group indicate configurations when the topology of the magnetic field can change.

As the magnetic field changes continuously in time, the position of fixed points and the elements of the matrix $\mm$ change continuously~\cite{smiet2019mapping}.
Exceptions to this occur when fixed points appear or disappear (always in pairs) or when a continuous line of fixed points breaks into an island chain~\cite{smiet2019mapping}.
$\mm$ traces a path through $\SL$ and continuous functions on the Lie group, e.g. $\trm$ and $q_0$, also change continuously.
There are two values of $\trm$ where $\mm$ can transition between the different subsets of $\SL$ and the topology of the field around the fixed point changes.
For example, the Kadomtsev process involves fixed points on the $q=1$ surface, with $\trm= +2$ and with $\mm$ constituting a shear mapping, which change into a hyperbolic fixed point, an X-point with $\trm\gtrsim2$, and an elliptic fixed point, the O-point of a $(1,1)$ island with $\trm\lesssim2$, which is created with a local value of $q_0=1$.
In the Kadomtsev model this point goes on to become the new axis.

When $\trm=-2$ an elliptic fixed point can become alternating-hyperbolic.
From Fig.~\ref{fig:sl2r} we can see that this transition can only occur when $q_0=2/(1+ 2n)$.
In this case, the field lines make an integer number \emph{and a half} rotations around the axis.
If this occurs at the magnetic axis, the axis becomes an alternating-hyperbolic X-point.
This can happen when $q_0=2/3$, which is very close to $q_0=0.7\pm0.1$, the value at which experiments have shown some crashes to be triggered~\cite{soltwisch1995sawtooth, levinton1993q}.

In order to explain the rapidity of the crash we require a fast instability that drives this transition.
We will now study this situation using ideal MHD theory, and we will present a calculation that illustrates that alternating-hyperbolic fixed points are prone to the formation of chaos.

We construct two families of MHD equilibria each with a circular-cross section, a minor radius equal to $1\mathrm{m}$, a major radius equal to $3\mathrm{m}$, a plasma $\beta_0$ of $3\%$ on the magnetic axis, and a toroidal field of $1\mathrm{T}$.
The pressure profile is quadratic in minor radius with a maximum on axis and zero on edge, and uniquely defined by the constraint that $\beta_0=3\%$ on the axis.
The first family of equilibria has a $q$-profile given by $q=q_0+\GERRITSCALE\psi_p$, where $\psi_p$ is the normalized poloidal flux function.
The safety factor on axis is thus $q_0$, and since $(r/a) \approx \sqrt{\psi_p}$, it increases quadratically with a value on the edge of $q_a=q_0+\GERRITSCALE \approx 3.8$.
These values are comparable to the parameters of typical of TFTR shots, which had a major radius $R=2.52\mathrm{m}$, minor radius $r=0.87\mathrm{m}$, and a plasma $\beta=1-8\%$ and a safety factor on edge $q_a=3-6$. The magnetic field is lower than on TFTR which was at maximum $6\mathrm{T}$.

We construct a second family of equilibria with a shallower safety factor profile, quartic in minor radius, given by $q = q_0 + 0.9 \psi_p^2$.
This profile has a low value of $q_a$ that would be challenging to achieve experimentally, but the shallow profile highlights modes that occur around the axis.
We investigate the ideal stability of these MHD equilibria with $q_0 \in [0.6,1.1]$ using the well-benchmarked NOVA-K code~\cite{cheng1992kinetic}.

We perform NOVA-K calculations in the ideal mode, not including kinetic effects.
The code finds ideal MHD modes in equilibria by solving the normal mode formulation of the linearized ideal MHD stability equations.
The normal mode equation is given by
   \begin{equation}
       -\omega^2 \rho \, \boldsymbol{\xi} = \bm{F}[\boldsymbol{\xi}],
   \end{equation}
   where $\omega^2$ is the mode frequency squared, $\rho$ is the plasma density, $\bm{F} $ is the linearized MHD force operator, and $\boldsymbol{\xi}$ is the displacement vector.
   Solutions with $\omega^2<0$ correspond to unstable, exponentially growing modes, and thus a displacement that exponentially increases.

\begin{figure}
       \includegraphics[width=8cm]{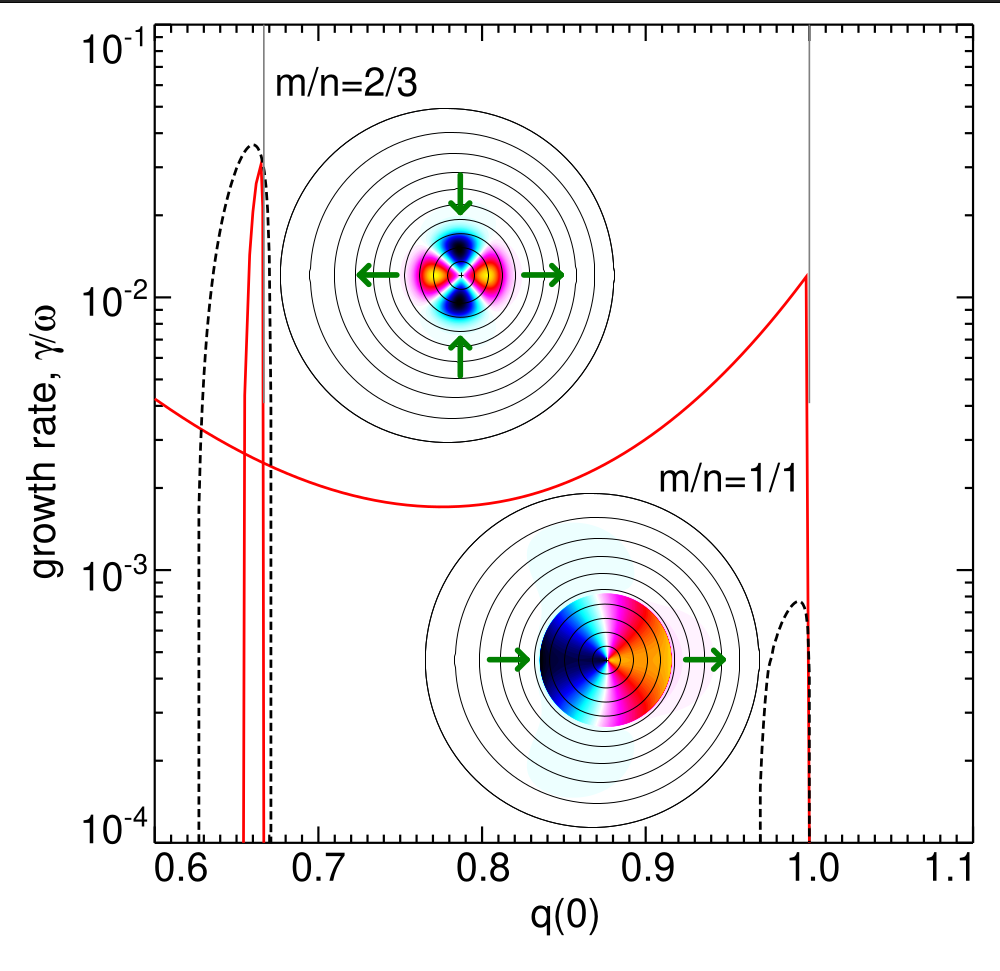}
       \caption{Ideal growth rates as a function of $q_0$ on the magnetic axis calculated by the NOVA-K code. The first family with $q=q_0+\GERRITSCALE\psi_p$ is given in red solid lines, and the second family with $q=q_0+0.9\psi_p^2$ in black dashed lines.
        Insets: radial component of the displacement vector of the ideal modes.
        The arrows indicate the direction of the displacement.}\label{fig:growthrates}
\end{figure}

In both families of equilibria we find two unstable ideal modes: a $1/1$ internal kink mode when $q_0 < 1$, and a $2/3$ mode when $q_0\sim2/3$ as shown in Fig.~\ref{fig:growthrates}.
In the first family, shown in red, we see an unstable 1/1 mode for the entire range $q_0<1$, as is expected from literature.
It is this mode that drives the Kadomtsev and Wesson processes, displacing the hot core within the $q=1$ surface.
In the second family of equilibria (black curves) the $q$-profile is much flatter and shallower, and the $1/1$ mode is stabilized by the boundary and the different shear at the $q=1$ surface.
In both families, exactly when $q_0$ reaches 2/3, a 2/3 mode appears with a much higher growth rate than the 1/1 mode
The radial component of the displacement of the two modes (taken from the second family of equilibria) is shown in the insets.

The $2/3$ mode is localized on the axis, and is present when an infinitesimal change to the field line map, caused by an infinitesimal change in the field, can change the mapping around the axis from elliptic to alternating-hyperbolic, as seen in figure~\ref{fig:sl2r}.
When $q_0=2/3$, field lines make one and a half rotations around the axis, and thus end up on the exact opposite side of the axis from whence they start.
A perturbation that causes this transition would be one for which a set of field lines (i.e. the slow manifold) is mapped closer to the fixed point (ending up opposite the axis but closer), and another set (i.e. the fast manifold) is mapped further away.
This is what the ideal displacement of a 2/3 mode accomplishes.
The displacement is directed towards the axis in two directions, and away from the axis in the two others, and rotating as a function of toroidal angle with the same rate as the field lines, such that a field line is resonantly displaced towards or away from the axis.

\begin{figure}  \includegraphics[width=8cm]{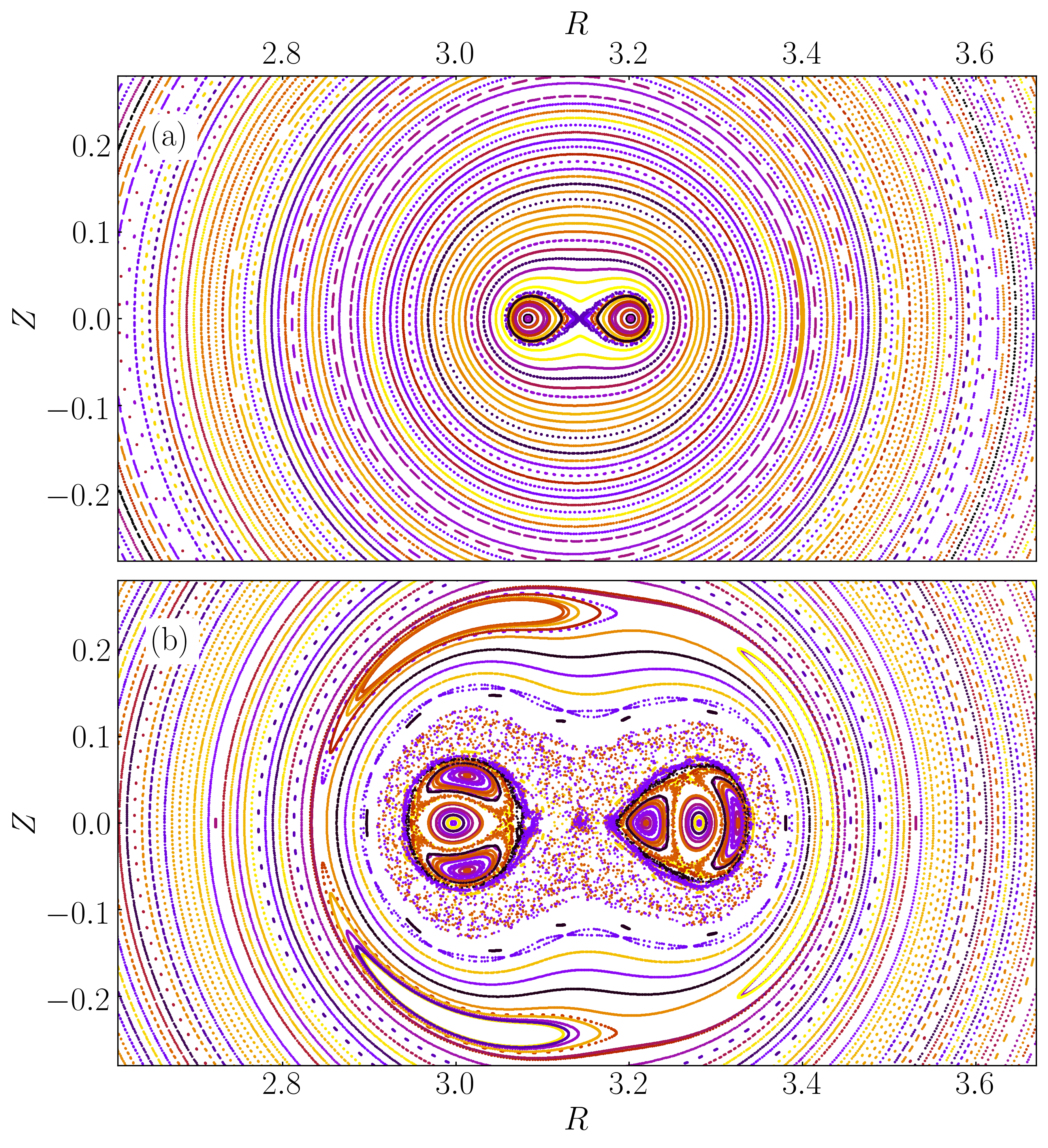}
  \caption{Effect of the $2/3$ displacement mode on the magnetic field of a tokamak equilibrium. (a): At a low perturbation amplitude of $A=4\times10^{-4}$ the magnetic topology around the axis changes into the alternating-hyperbolic configuration. (b): At a higher amplitude $A=0.01$ a finite region around the core stochastisizes.}
\label{fig:perturbed}
\end{figure}

To demonstrate this we will show the effect of an analytic 2/3 mode on the magnetic topology by perturbing a tokamak equilibrium field.
We construct the equilibrium with a quadratic safety factor profile given by $q=2/3 + 2.3333*\psi_p$, and all other parameters identical to the families of equilibria described above.
The analytic displacement is calculated from:
\begin{equation}
  \Psi=A\exp\left(-\frac{\psi_p}{\sigma}\right)\cos(2\theta-3\phi)
\end{equation}
where $\theta$ is the poloidal angle, $A$ is an overall scaling factor, and the radial envelope has a characteristic width $\sigma=0.15$.
The displacement vector $\boldsymbol{\xi}$ is calculated through $\xi_R=-(1/R) \partial_Z\Psi$ and $\xi_Z=(1/R) \partial_R\Psi$.
In the normal mode formulation of the ideal MHD equations, the perturbed field that corresponds with a displacement is given by
\begin{equation}
\delta\bm{B} = \nabla\times \left(\boldsymbol{\xi}\times\bm{B_0}\right)
\end{equation}
where $\bm{B}_0$ is the equilibrium magnetic field.

We analyze the structure of the magnetic field by calculating the trajectories of low-energy ($1 \mathrm{keV}$) electrons using the SPIRAL code~\cite{kramer2013description}.
The drift orbit of these low-energy electrons is so small that they are effectively field-line following.
We calculate the Poincar\'e map 1000 times per trajectory for 127 starting points equidistantly spaced between the equilibrium axis and the plasma boundary.

Poincar\'e plots of $\bm{B_0}+\delta\bm{B}$ are shown in Fig.~\ref{fig:perturbed}.
The scaling factor $A=4\times10^{-4}$ results in a perturbed field $|\delta \bm{B}|/|\bm{B}_{0}| \sim 1\times 10^{-3}$.
At this amplitude the nested flux surfaces near the axis are broken up through resonance with the applied perturbation as shown in Fig.~\ref{fig:perturbed}(a).
Note that the center of the two `islands' are themselves not fixed points of the field line map: the one maps to the other.
The axis is an alternating-hyperbolic fixed point.

At a value of $A=0.01$, which corresponds to $|\delta \bm{B}|/|\bm{B}_{0}| \sim 2.5\times 10^{-2}$, the field around the magnetic axis becomes chaotic, shown in Fig.~\ref{fig:perturbed}(b).
A 3/3 island chain is seen at the $q=1$ surface.
The $1/1$ surface is not removed by this perturbation.

Although a nonlinear calculation is beyond the scope of this paper, it is possible to conjecture what effects the observed change in magnetic topology will have on the equilibrium.
At small amplitudes, the change in topology magnetically connects the plasma at the magnetic axis with further out.
Since the plasma pressure is highest on the axis, this causes a flow along the field lines.
This flow can excite other modes and further perturb the field.
The core region stochastisizes and becomes magnetically connected.
The disappearance of stabilizing nested flux surfaces can trigger other modes, notably the 1/1 mode which can still be unstable.
This could explain why soft X-ray measurements of sawteeth where $q_0$ is measured to be 0.7 still often show a 1/1 temperature distribution during the crash phase~\cite{nagayama1996tomography, yamada1994investigation}.
Stochastisation happens for a pure $2/3$ mode at the amplitude show in Fig.~\ref{fig:perturbed}(b), but can occur more readily if other modes are involved.
The temperature and pressure are equilibrated in such a connected region through rapid parallel transport.
Stochastisation of a magnetic field is known to drive rapid reconnection~\cite{boozer2014formation, huang2014rapid}, which redistributes the poloidal and toroidal fluxes within the stochastic region.

The above considerations lead us to formulate the alternating-hyperbolic sawtooth model:
Because of current diffusion, $q_0$ decreases on a slow, resistive timescale.
When the $1/1$ internal kink is stabilized, thereby preventing a crash that resets $q_0$ to $1$, the safety factor decreases to near $q_0=2/3$.
The ideal $2/3$ mode causes the axis to transition into an alternating-hyperbolic fixed point and rapidly stochastisizes a region around the axis.
This can trigger other modes (further increasing stochasticity) such as the unstable 1/1 mode which is visible in temperature diagnostics.
The stochastic region can be restricted to within the $q=1$ surface (consistent with an inversion radius smaller than the radius of the $q=1$ surface~\cite{rice1992fifteen, soltwisch1995sawtooth}), leaving that surface intact during the crash.
The poloidal and toroidal fluxes are redistributed, resulting in an increased safety factor.
This shifts the field out of resonance with the mode and the core region heals with $2/3<q_0<1$.
After the crash, current diffusion again slowly decreases the safety factor until $q_0=2/3$ is reached, and the crash occurs anew.

This model explains why all the observations of low-$q$ sawteeth occur clustered around the value of $q=0.7\pm0.1$~\cite{west1987measurement, osborne1982discharges, wolf1993comparison, rice1992fifteen, soltwisch1986current, soltwisch1988measurement, soltwisch1995sawtooth, levinton1993q, levinton1993q, yamada1994investigation}; A crash triggered at $q_0=2/3$ lies squarely within the experimental uncertainty of all these measurements.
A snake~\cite{weller1987persistent}, can thus remain as the $q=1$ surface is unaffected by this crash.
The same holds for EAEs~\cite{kramer2001magnetic} that live on the $q=1$ surface and are present immediately after the crash.

One may wonder how the tokamak reaches a state with $q_0$ this significantly below $1$, but direct measurements tell us it can~\cite{west1987measurement, osborne1982discharges, rice1992fifteen, soltwisch1986current, soltwisch1988measurement, soltwisch1995sawtooth, levinton1993q, levinton1993q, yamada1994investigation}
There are several reported mechanisms that stabilize the internal kink mode~\cite{chapman2007physics}, including toroidal rotation~\cite{chapman2006effect}, fast particles~\cite{cole2014fluid, porcelli1991fast}, and diamagnetic effects~\cite{connor2012unified}.
Stabilization is experimentally confirmed by the decrease of sawtooth repetition rate through fast particle injection~\cite{angioni2002neutral, campbell1988stabilization}.
(In such stabilized regimes, sawteeth with very low repetition rate can also be observed, so-called giant sawteeth, \cite{bernabei2001combined, campbell1988stabilization}, which do exhibit a re-set to $q_0=1$.
These can be triggered by the onset of energetic particle modes that remove the stabilizing fast particles~\cite{bernabei2001combined}.)

In this paper we have focused on the sawtooth events that occur when $q_0=2/3$, but as figure~\ref{fig:sl2r} shows, a transition of the axis to a(n) (alternating-)hyperbolic point can occur when $q=1/n$,  (respectively $q=1/(1/2+n)$).
The new explanation of the sawtooth proposed by Jardin et al.~\cite{jardin2020new} posits that crashes occur in a discharge where $q_0$ is clamped just above 1 through a nonlinearly saturated quasi-interchange flow, and are triggered by crossing the threshold of the $2/2$ mode.
The 2/2 mode induces the same displacement towards and away from the axis, which when $q_0=1$ would drive the transition to a (regular) hyperbolic geometry.
The presented numerical simulation (see fig. 6 (b)) exhibits a stochastic core region surrounding a 2/2 structure similar to figure\ref{fig:perturbed}(b)  which causes the crash.
Sawtooth-like relaxation events can also be observed when $q=2$~\cite{chang1996off} and are traditionally attributed to a 2/1 double tearing instability.
The present work suggests that a transition to a(n) (alternating-)hyperbolic transition could form a unified theory that explains all three phenomena.

The alternating-hyperbolic sawtooth model can possibly be measured on tokamaks with sufficiently fast diagnostics.
SRX tomography (f.ex.~\cite{nagayama1996tomography}) shows that the crash phase is often (but not always) dominated by a 1/1 signature.
This would suggest that the 2/3 mode can act as a very fast trigger, which in turn sets off the 1/1 mode.
The SRX data  in~\cite{nagayama1996tomography} does show a few chords with a faster oscillation frequency just before temperature equilibration is initiated, but this instant is not tomographically reconstructed.
On a modern tokamak electron cyclotron emission imaging (ECEI) would be fast enough to discern the structure of the mode that initiates the crash.

Most measurements of $q_0\sim0.7$ have been performed on circular cross-section discharges, and after devices switched to an elliptic or D-shaped geometry, we mostly see crashes that occur when $q\sim1$ (see f.ex. recent observations on KSTAR~\cite{nam2018validation}).
The measurements of $q_0=0.7$, which have long stood in the way of a full understanding of the ubiquitous sawtooth phenomenon, can be explained by this alternating-hyperbolic model.

\begin{acknowledgments}
CBS would like to thank Wei Zhang for sharing data on the $q=2/1$ double tearing crash that planted the idea for this 2/3 model.
We would like to thank Stephen Jardin, Hyeon Park, and Amitava Bhattacharjee for helpful discussions.
CBS acknowledges support from the Rubicon programme with project number 680-50-1532, which is (partly)
financed by the Netherlands Organization for Scientific Research (NWO).
Notice: This manuscript is based upon work supported by the U.S. Department of Energy, Office of Science, Office of Fusion Energy Sciences, and has been authored by Princeton University under Contract Number DE-AC02-09CH11466 with the U.S. Department of Energy. The publisher, by accepting the article for publication acknowledges, that the United States Government retains a non-exclusive, paid-up, irrevocable, world-wide license to publish or reproduce the published form of this manuscript, or allow others to do so, for United States Government purposes.
\end{acknowledgments}


\begin{thebibliography}{47}%
\makeatletter
\providecommand \@ifxundefined [1]{%
 \@ifx{#1\undefined}
}%
\providecommand \@ifnum [1]{%
 \ifnum #1\expandafter \@firstoftwo
 \else \expandafter \@secondoftwo
 \fi
}%
\providecommand \@ifx [1]{%
 \ifx #1\expandafter \@firstoftwo
 \else \expandafter \@secondoftwo
 \fi
}%
\providecommand \natexlab [1]{#1}%
\providecommand \enquote  [1]{``#1''}%
\providecommand \bibnamefont  [1]{#1}%
\providecommand \bibfnamefont [1]{#1}%
\providecommand \citenamefont [1]{#1}%
\providecommand \href@noop [0]{\@secondoftwo}%
\providecommand \href [0]{\begingroup \@sanitize@url \@href}%
\providecommand \@href[1]{\@@startlink{#1}\@@href}%
\providecommand \@@href[1]{\endgroup#1\@@endlink}%
\providecommand \@sanitize@url [0]{\catcode `\\12\catcode `\$12\catcode
  `\&12\catcode `\#12\catcode `\^12\catcode `\_12\catcode `\%12\relax}%
\providecommand \@@startlink[1]{}%
\providecommand \@@endlink[0]{}%
\providecommand \url  [0]{\begingroup\@sanitize@url \@url }%
\providecommand \@url [1]{\endgroup\@href {#1}{\urlprefix }}%
\providecommand \urlprefix  [0]{URL }%
\providecommand \Eprint [0]{\href }%
\providecommand \doibase [0]{http://dx.doi.org/}%
\providecommand \selectlanguage [0]{\@gobble}%
\providecommand \bibinfo  [0]{\@secondoftwo}%
\providecommand \bibfield  [0]{\@secondoftwo}%
\providecommand \translation [1]{[#1]}%
\providecommand \BibitemOpen [0]{}%
\providecommand \bibitemStop [0]{}%
\providecommand \bibitemNoStop [0]{.\EOS\space}%
\providecommand \EOS [0]{\spacefactor3000\relax}%
\providecommand \BibitemShut  [1]{\csname bibitem#1\endcsname}%
\let\auto@bib@innerbib\@empty
\bibitem [{\citenamefont {Von~Goeler}\ \emph {et~al.}(1974)\citenamefont
  {Von~Goeler}, \citenamefont {Stodiek},\ and\ \citenamefont
  {Sauthoff}}]{von1974studies}%
  \BibitemOpen
  \bibfield  {author} {\bibinfo {author} {\bibfnamefont {S.}~\bibnamefont
  {Von~Goeler}}, \bibinfo {author} {\bibfnamefont {W.}~\bibnamefont {Stodiek}},
  \ and\ \bibinfo {author} {\bibfnamefont {N.}~\bibnamefont {Sauthoff}},\
  }\href@noop {} {\bibfield  {journal} {\bibinfo  {journal} {Physical Review
  Letters}\ }\textbf {\bibinfo {volume} {33}},\ \bibinfo {pages} {1201}
  (\bibinfo {year} {1974})}\BibitemShut {NoStop}%
\bibitem [{\citenamefont {Chapman}(2010)}]{chapman2010controlling}%
  \BibitemOpen
  \bibfield  {author} {\bibinfo {author} {\bibfnamefont {I.~T.}\ \bibnamefont
  {Chapman}},\ }\href@noop {} {\bibfield  {journal} {\bibinfo  {journal}
  {Plasma Physics and Controlled Fusion}\ }\textbf {\bibinfo {volume} {53}},\
  \bibinfo {pages} {013001} (\bibinfo {year} {2010})}\BibitemShut {NoStop}%
\bibitem [{\citenamefont {Kadomtsev}(1975)}]{kadomtsev1975disruptive}%
  \BibitemOpen
  \bibfield  {author} {\bibinfo {author} {\bibfnamefont {B.~B.}\ \bibnamefont
  {Kadomtsev}},\ }\href@noop {} {\bibfield  {journal} {\bibinfo  {journal}
  {Soviet Journal of Plasma Physics}\ }\textbf {\bibinfo {volume} {1}},\
  \bibinfo {pages} {710} (\bibinfo {year} {1975})}\BibitemShut {NoStop}%
\bibitem [{\citenamefont {Coppi}\ \emph {et~al.}(1976)\citenamefont {Coppi},
  \citenamefont {Galvao}, \citenamefont {Pellat}, \citenamefont {Rosenbluth},\
  and\ \citenamefont {Rutherford}}]{coppi1976resistive}%
  \BibitemOpen
  \bibfield  {author} {\bibinfo {author} {\bibfnamefont {B.}~\bibnamefont
  {Coppi}}, \bibinfo {author} {\bibfnamefont {R.}~\bibnamefont {Galvao}},
  \bibinfo {author} {\bibfnamefont {R.}~\bibnamefont {Pellat}}, \bibinfo
  {author} {\bibfnamefont {M.}~\bibnamefont {Rosenbluth}}, \ and\ \bibinfo
  {author} {\bibfnamefont {P.}~\bibnamefont {Rutherford}},\ }\href@noop {}
  {\bibfield  {journal} {\bibinfo  {journal} {Fizika Plazmy}\ }\textbf
  {\bibinfo {volume} {2}},\ \bibinfo {pages} {961} (\bibinfo {year}
  {1976})}\BibitemShut {NoStop}%
\bibitem [{\citenamefont {Wesson}(1986)}]{wesson1986sawtooth}%
  \BibitemOpen
  \bibfield  {author} {\bibinfo {author} {\bibfnamefont {J.~A.}\ \bibnamefont
  {Wesson}},\ }\href@noop {} {\bibfield  {journal} {\bibinfo  {journal} {Plasma
  physics and controlled fusion}\ }\textbf {\bibinfo {volume} {28}},\ \bibinfo
  {pages} {243} (\bibinfo {year} {1986})}\BibitemShut {NoStop}%
\bibitem [{\citenamefont {Jardin}\ \emph {et~al.}(2015)\citenamefont {Jardin},
  \citenamefont {Ferraro},\ and\ \citenamefont {Krebs}}]{jardin2015self}%
  \BibitemOpen
  \bibfield  {author} {\bibinfo {author} {\bibfnamefont {S.~C.}\ \bibnamefont
  {Jardin}}, \bibinfo {author} {\bibfnamefont {N.}~\bibnamefont {Ferraro}}, \
  and\ \bibinfo {author} {\bibfnamefont {I.}~\bibnamefont {Krebs}},\
  }\href@noop {} {\bibfield  {journal} {\bibinfo  {journal} {Physical review
  letters}\ }\textbf {\bibinfo {volume} {115}},\ \bibinfo {pages} {215001}
  (\bibinfo {year} {2015})}\BibitemShut {NoStop}%
\bibitem [{\citenamefont {Jardin}\ \emph {et~al.}(2020)\citenamefont {Jardin},
  \citenamefont {Krebs},\ and\ \citenamefont {Ferraro}}]{jardin2020new}%
  \BibitemOpen
  \bibfield  {author} {\bibinfo {author} {\bibfnamefont {S.}~\bibnamefont
  {Jardin}}, \bibinfo {author} {\bibfnamefont {I.}~\bibnamefont {Krebs}}, \
  and\ \bibinfo {author} {\bibfnamefont {N.}~\bibnamefont {Ferraro}},\
  }\href@noop {} {\bibfield  {journal} {\bibinfo  {journal} {Physics of
  Plasmas}\ }\textbf {\bibinfo {volume} {27}},\ \bibinfo {pages} {032509}
  (\bibinfo {year} {2020})}\BibitemShut {NoStop}%
\bibitem [{\citenamefont {Weller}\ \emph {et~al.}(1987)\citenamefont {Weller},
  \citenamefont {Cheetham}, \citenamefont {Edwards}, \citenamefont {Gill},
  \citenamefont {Gondhalekar}, \citenamefont {Granetz}, \citenamefont
  {Snipes},\ and\ \citenamefont {Wesson}}]{weller1987persistent}%
  \BibitemOpen
  \bibfield  {author} {\bibinfo {author} {\bibfnamefont {A.}~\bibnamefont
  {Weller}}, \bibinfo {author} {\bibfnamefont {A.~D.}\ \bibnamefont
  {Cheetham}}, \bibinfo {author} {\bibfnamefont {A.~W.}\ \bibnamefont
  {Edwards}}, \bibinfo {author} {\bibfnamefont {R.~D.}\ \bibnamefont {Gill}},
  \bibinfo {author} {\bibfnamefont {A.}~\bibnamefont {Gondhalekar}}, \bibinfo
  {author} {\bibfnamefont {R.~S.}\ \bibnamefont {Granetz}}, \bibinfo {author}
  {\bibfnamefont {J.}~\bibnamefont {Snipes}}, \ and\ \bibinfo {author}
  {\bibfnamefont {J.~A.}\ \bibnamefont {Wesson}},\ }\href@noop {} {\bibfield
  {journal} {\bibinfo  {journal} {Physical review letters}\ }\textbf {\bibinfo
  {volume} {59}},\ \bibinfo {pages} {2303} (\bibinfo {year}
  {1987})}\BibitemShut {NoStop}%
\bibitem [{\citenamefont {Wroblewski}\ and\ \citenamefont
  {Lao}(1991)}]{wroblewski1991determination}%
  \BibitemOpen
  \bibfield  {author} {\bibinfo {author} {\bibfnamefont {D.}~\bibnamefont
  {Wroblewski}}\ and\ \bibinfo {author} {\bibfnamefont {L.}~\bibnamefont
  {Lao}},\ }\href@noop {} {\bibfield  {journal} {\bibinfo  {journal} {Physics
  of Fluids B: Plasma Physics}\ }\textbf {\bibinfo {volume} {3}},\ \bibinfo
  {pages} {2877} (\bibinfo {year} {1991})}\BibitemShut {NoStop}%
\bibitem [{\citenamefont {Nam}\ \emph {et~al.}(2018)\citenamefont {Nam},
  \citenamefont {Ko}, \citenamefont {Choe}, \citenamefont {Bae}, \citenamefont
  {Choi}, \citenamefont {Lee}, \citenamefont {Yun}, \citenamefont {Jardin},\
  and\ \citenamefont {Park}}]{nam2018validation}%
  \BibitemOpen
  \bibfield  {author} {\bibinfo {author} {\bibfnamefont {Y.~B.}\ \bibnamefont
  {Nam}}, \bibinfo {author} {\bibfnamefont {J.~S.}\ \bibnamefont {Ko}},
  \bibinfo {author} {\bibfnamefont {G.~H.}\ \bibnamefont {Choe}}, \bibinfo
  {author} {\bibfnamefont {Y.}~\bibnamefont {Bae}}, \bibinfo {author}
  {\bibfnamefont {M.~J.}\ \bibnamefont {Choi}}, \bibinfo {author}
  {\bibfnamefont {W.}~\bibnamefont {Lee}}, \bibinfo {author} {\bibfnamefont
  {G.~S.}\ \bibnamefont {Yun}}, \bibinfo {author} {\bibfnamefont
  {S.}~\bibnamefont {Jardin}}, \ and\ \bibinfo {author} {\bibfnamefont
  {H.}~\bibnamefont {Park}},\ }\href@noop {} {\bibfield  {journal} {\bibinfo
  {journal} {Nuclear Fusion}\ }\textbf {\bibinfo {volume} {58}},\ \bibinfo
  {pages} {066009} (\bibinfo {year} {2018})}\BibitemShut {NoStop}%
\bibitem [{\citenamefont {West}\ \emph {et~al.}(1987)\citenamefont {West},
  \citenamefont {Thomas}, \citenamefont {DeGrassie},\ and\ \citenamefont
  {Zheng}}]{west1987measurement}%
  \BibitemOpen
  \bibfield  {author} {\bibinfo {author} {\bibfnamefont {W.}~\bibnamefont
  {West}}, \bibinfo {author} {\bibfnamefont {D.}~\bibnamefont {Thomas}},
  \bibinfo {author} {\bibfnamefont {J.}~\bibnamefont {DeGrassie}}, \ and\
  \bibinfo {author} {\bibfnamefont {S.}~\bibnamefont {Zheng}},\ }\href@noop {}
  {\bibfield  {journal} {\bibinfo  {journal} {Physical review letters}\
  }\textbf {\bibinfo {volume} {58}},\ \bibinfo {pages} {2758} (\bibinfo {year}
  {1987})}\BibitemShut {NoStop}%
\bibitem [{\citenamefont {Osborne}\ \emph {et~al.}(1982)\citenamefont
  {Osborne}, \citenamefont {Dexter},\ and\ \citenamefont
  {Prager}}]{osborne1982discharges}%
  \BibitemOpen
  \bibfield  {author} {\bibinfo {author} {\bibfnamefont {T.}~\bibnamefont
  {Osborne}}, \bibinfo {author} {\bibfnamefont {R.}~\bibnamefont {Dexter}}, \
  and\ \bibinfo {author} {\bibfnamefont {S.~C.}\ \bibnamefont {Prager}},\
  }\href@noop {} {\bibfield  {journal} {\bibinfo  {journal} {Physical Review
  Letters}\ }\textbf {\bibinfo {volume} {49}},\ \bibinfo {pages} {734}
  (\bibinfo {year} {1982})}\BibitemShut {NoStop}%
\bibitem [{\citenamefont {Rice}\ and\ \citenamefont
  {Hooper}(1994)}]{rice1994poloidal}%
  \BibitemOpen
  \bibfield  {author} {\bibinfo {author} {\bibfnamefont {B.}~\bibnamefont
  {Rice}}\ and\ \bibinfo {author} {\bibfnamefont {E.}~\bibnamefont {Hooper}},\
  }\href@noop {} {\bibfield  {journal} {\bibinfo  {journal} {Nuclear fusion}\
  }\textbf {\bibinfo {volume} {34}},\ \bibinfo {pages} {1} (\bibinfo {year}
  {1994})}\BibitemShut {NoStop}%
\bibitem [{\citenamefont {Soltwisch}(1986)}]{soltwisch1986current}%
  \BibitemOpen
  \bibfield  {author} {\bibinfo {author} {\bibfnamefont {H.}~\bibnamefont
  {Soltwisch}},\ }\href@noop {} {\bibfield  {journal} {\bibinfo  {journal}
  {Review of Scientific Instruments}\ }\textbf {\bibinfo {volume} {57}},\
  \bibinfo {pages} {1939} (\bibinfo {year} {1986})}\BibitemShut {NoStop}%
\bibitem [{\citenamefont {Soltwisch}(1988)}]{soltwisch1988measurement}%
  \BibitemOpen
  \bibfield  {author} {\bibinfo {author} {\bibfnamefont {H.}~\bibnamefont
  {Soltwisch}},\ }\href@noop {} {\bibfield  {journal} {\bibinfo  {journal}
  {Review of Scientific Instruments}\ }\textbf {\bibinfo {volume} {59}},\
  \bibinfo {pages} {1599} (\bibinfo {year} {1988})}\BibitemShut {NoStop}%
\bibitem [{\citenamefont {Soltwisch}\ and\ \citenamefont
  {Koslowski}(1995)}]{soltwisch1995sawtooth}%
  \BibitemOpen
  \bibfield  {author} {\bibinfo {author} {\bibfnamefont {H.}~\bibnamefont
  {Soltwisch}}\ and\ \bibinfo {author} {\bibfnamefont {H.}~\bibnamefont
  {Koslowski}},\ }\href@noop {} {\bibfield  {journal} {\bibinfo  {journal}
  {Plasma physics and controlled fusion}\ }\textbf {\bibinfo {volume} {37}},\
  \bibinfo {pages} {667} (\bibinfo {year} {1995})}\BibitemShut {NoStop}%
\bibitem [{\citenamefont {Levinton}\ \emph {et~al.}(1993)\citenamefont
  {Levinton}, \citenamefont {Batha}, \citenamefont {Yamada},\ and\
  \citenamefont {Zarnstorff}}]{levinton1993q}%
  \BibitemOpen
  \bibfield  {author} {\bibinfo {author} {\bibfnamefont {F.~M.}\ \bibnamefont
  {Levinton}}, \bibinfo {author} {\bibfnamefont {S.~H.}\ \bibnamefont {Batha}},
  \bibinfo {author} {\bibfnamefont {M.}~\bibnamefont {Yamada}}, \ and\ \bibinfo
  {author} {\bibfnamefont {M.}~\bibnamefont {Zarnstorff}},\ }\href@noop {}
  {\bibfield  {journal} {\bibinfo  {journal} {Physics of Fluids B: Plasma
  Physics}\ }\textbf {\bibinfo {volume} {5}},\ \bibinfo {pages} {2554}
  (\bibinfo {year} {1993})}\BibitemShut {NoStop}%
\bibitem [{\citenamefont {Yamada}\ \emph {et~al.}(1994)\citenamefont {Yamada},
  \citenamefont {Levinton}, \citenamefont {Pomphrey}, \citenamefont {Budny},
  \citenamefont {Manickam},\ and\ \citenamefont
  {Nagayama}}]{yamada1994investigation}%
  \BibitemOpen
  \bibfield  {author} {\bibinfo {author} {\bibfnamefont {M.}~\bibnamefont
  {Yamada}}, \bibinfo {author} {\bibfnamefont {F.}~\bibnamefont {Levinton}},
  \bibinfo {author} {\bibfnamefont {N.}~\bibnamefont {Pomphrey}}, \bibinfo
  {author} {\bibfnamefont {R.}~\bibnamefont {Budny}}, \bibinfo {author}
  {\bibfnamefont {J.}~\bibnamefont {Manickam}}, \ and\ \bibinfo {author}
  {\bibfnamefont {Y.}~\bibnamefont {Nagayama}},\ }\href@noop {} {\bibfield
  {journal} {\bibinfo  {journal} {Physics of plasmas}\ }\textbf {\bibinfo
  {volume} {1}},\ \bibinfo {pages} {3269} (\bibinfo {year} {1994})}\BibitemShut
  {NoStop}%
\bibitem [{\citenamefont {Wolf}\ \emph {et~al.}(1993)\citenamefont {Wolf},
  \citenamefont {O'Rourke}, \citenamefont {Edwards},\ and\ \citenamefont
  {Von~Hellermann}}]{wolf1993comparison}%
  \BibitemOpen
  \bibfield  {author} {\bibinfo {author} {\bibfnamefont {R.}~\bibnamefont
  {Wolf}}, \bibinfo {author} {\bibfnamefont {J.}~\bibnamefont {O'Rourke}},
  \bibinfo {author} {\bibfnamefont {A.}~\bibnamefont {Edwards}}, \ and\
  \bibinfo {author} {\bibfnamefont {M.}~\bibnamefont {Von~Hellermann}},\
  }\href@noop {} {\bibfield  {journal} {\bibinfo  {journal} {Nuclear fusion}\
  }\textbf {\bibinfo {volume} {33}},\ \bibinfo {pages} {663} (\bibinfo {year}
  {1993})}\BibitemShut {NoStop}%
\bibitem [{\citenamefont {Rice}(1992)}]{rice1992fifteen}%
  \BibitemOpen
  \bibfield  {author} {\bibinfo {author} {\bibfnamefont {B.}~\bibnamefont
  {Rice}},\ }\href@noop {} {\bibfield  {journal} {\bibinfo  {journal} {Review
  of scientific instruments}\ }\textbf {\bibinfo {volume} {63}},\ \bibinfo
  {pages} {5002} (\bibinfo {year} {1992})}\BibitemShut {NoStop}%
\bibitem [{\citenamefont {Park}(2019)}]{park2019newly}%
  \BibitemOpen
  \bibfield  {author} {\bibinfo {author} {\bibfnamefont {H.~K.}\ \bibnamefont
  {Park}},\ }\href@noop {} {\bibfield  {journal} {\bibinfo  {journal} {Advances
  in Physics: X}\ }\textbf {\bibinfo {volume} {4}},\ \bibinfo {pages} {1633956}
  (\bibinfo {year} {2019})}\BibitemShut {NoStop}%
\bibitem [{\citenamefont {Gill}\ \emph {et~al.}(1992)\citenamefont {Gill},
  \citenamefont {Edwards}, \citenamefont {Pasini},\ and\ \citenamefont
  {Weller}}]{gill1992snake}%
  \BibitemOpen
  \bibfield  {author} {\bibinfo {author} {\bibfnamefont {R.~D.}\ \bibnamefont
  {Gill}}, \bibinfo {author} {\bibfnamefont {A.~W.}\ \bibnamefont {Edwards}},
  \bibinfo {author} {\bibfnamefont {D.}~\bibnamefont {Pasini}}, \ and\ \bibinfo
  {author} {\bibfnamefont {A.}~\bibnamefont {Weller}},\ }\href@noop {}
  {\bibfield  {journal} {\bibinfo  {journal} {Nuclear Fusion}\ }\textbf
  {\bibinfo {volume} {32}},\ \bibinfo {pages} {723} (\bibinfo {year}
  {1992})}\BibitemShut {NoStop}%
\bibitem [{\citenamefont {Biskamp}\ and\ \citenamefont
  {Drake}(1994)}]{biskamp1994dynamics}%
  \BibitemOpen
  \bibfield  {author} {\bibinfo {author} {\bibfnamefont {D.}~\bibnamefont
  {Biskamp}}\ and\ \bibinfo {author} {\bibfnamefont {J.}~\bibnamefont
  {Drake}},\ }\href@noop {} {\bibfield  {journal} {\bibinfo  {journal}
  {Physical review letters}\ }\textbf {\bibinfo {volume} {73}},\ \bibinfo
  {pages} {971} (\bibinfo {year} {1994})}\BibitemShut {NoStop}%
\bibitem [{\citenamefont {Kramer}\ \emph {et~al.}(2001)\citenamefont {Kramer},
  \citenamefont {Cheng}, \citenamefont {Kusama}, \citenamefont {Nazikian},
  \citenamefont {Takeji},\ and\ \citenamefont {Tobita}}]{kramer2001magnetic}%
  \BibitemOpen
  \bibfield  {author} {\bibinfo {author} {\bibfnamefont {G.~J.}\ \bibnamefont
  {Kramer}}, \bibinfo {author} {\bibfnamefont {C.~Z.}\ \bibnamefont {Cheng}},
  \bibinfo {author} {\bibfnamefont {Y.}~\bibnamefont {Kusama}}, \bibinfo
  {author} {\bibfnamefont {R.}~\bibnamefont {Nazikian}}, \bibinfo {author}
  {\bibfnamefont {S.}~\bibnamefont {Takeji}}, \ and\ \bibinfo {author}
  {\bibfnamefont {K.}~\bibnamefont {Tobita}},\ }\href@noop {} {\bibfield
  {journal} {\bibinfo  {journal} {Nuclear fusion}\ }\textbf {\bibinfo {volume}
  {41}},\ \bibinfo {pages} {1135} (\bibinfo {year} {2001})}\BibitemShut
  {NoStop}%
\bibitem [{\citenamefont {Beidler}\ and\ \citenamefont
  {Cassak}(2011)}]{beidler2011model}%
  \BibitemOpen
  \bibfield  {author} {\bibinfo {author} {\bibfnamefont {M.~T.}\ \bibnamefont
  {Beidler}}\ and\ \bibinfo {author} {\bibfnamefont {P.~A.}\ \bibnamefont
  {Cassak}},\ }\href@noop {} {\bibfield  {journal} {\bibinfo  {journal}
  {Physical review letters}\ }\textbf {\bibinfo {volume} {107}},\ \bibinfo
  {pages} {255002} (\bibinfo {year} {2011})}\BibitemShut {NoStop}%
\bibitem [{\citenamefont {Furno}\ \emph {et~al.}(2001)\citenamefont {Furno},
  \citenamefont {Angioni}, \citenamefont {Porcelli}, \citenamefont {Weisen},
  \citenamefont {Behn}, \citenamefont {Goodman}, \citenamefont {Henderson},
  \citenamefont {Pietrzyk}, \citenamefont {Pochelon}, \citenamefont {Reimerdes}
  \emph {et~al.}}]{furno2001understanding}%
  \BibitemOpen
  \bibfield  {author} {\bibinfo {author} {\bibfnamefont {I.}~\bibnamefont
  {Furno}}, \bibinfo {author} {\bibfnamefont {C.}~\bibnamefont {Angioni}},
  \bibinfo {author} {\bibfnamefont {F.}~\bibnamefont {Porcelli}}, \bibinfo
  {author} {\bibfnamefont {H.}~\bibnamefont {Weisen}}, \bibinfo {author}
  {\bibfnamefont {R.}~\bibnamefont {Behn}}, \bibinfo {author} {\bibfnamefont
  {T.}~\bibnamefont {Goodman}}, \bibinfo {author} {\bibfnamefont
  {M.}~\bibnamefont {Henderson}}, \bibinfo {author} {\bibfnamefont
  {Z.}~\bibnamefont {Pietrzyk}}, \bibinfo {author} {\bibfnamefont
  {A.}~\bibnamefont {Pochelon}}, \bibinfo {author} {\bibfnamefont
  {H.}~\bibnamefont {Reimerdes}},  \emph {et~al.},\ }\href@noop {} {\bibfield
  {journal} {\bibinfo  {journal} {Nuclear Fusion}\ }\textbf {\bibinfo {volume}
  {41}},\ \bibinfo {pages} {403} (\bibinfo {year} {2001})}\BibitemShut
  {NoStop}%
\bibitem [{\citenamefont {Pietrzyk}\ \emph {et~al.}(1999)\citenamefont
  {Pietrzyk}, \citenamefont {Pochelon}, \citenamefont {Goodman}, \citenamefont
  {Henderson}, \citenamefont {Hogge}, \citenamefont {Reimerdes}, \citenamefont
  {Tran}, \citenamefont {Behn}, \citenamefont {Furno}, \citenamefont {Moret}
  \emph {et~al.}}]{pietrzyk1999behaviour}%
  \BibitemOpen
  \bibfield  {author} {\bibinfo {author} {\bibfnamefont {Z.}~\bibnamefont
  {Pietrzyk}}, \bibinfo {author} {\bibfnamefont {A.}~\bibnamefont {Pochelon}},
  \bibinfo {author} {\bibfnamefont {T.}~\bibnamefont {Goodman}}, \bibinfo
  {author} {\bibfnamefont {M.}~\bibnamefont {Henderson}}, \bibinfo {author}
  {\bibfnamefont {J.-P.}\ \bibnamefont {Hogge}}, \bibinfo {author}
  {\bibfnamefont {H.}~\bibnamefont {Reimerdes}}, \bibinfo {author}
  {\bibfnamefont {M.}~\bibnamefont {Tran}}, \bibinfo {author} {\bibfnamefont
  {R.}~\bibnamefont {Behn}}, \bibinfo {author} {\bibfnamefont {I.}~\bibnamefont
  {Furno}}, \bibinfo {author} {\bibfnamefont {J.-M.}\ \bibnamefont {Moret}},
  \emph {et~al.},\ }\href@noop {} {\bibfield  {journal} {\bibinfo  {journal}
  {Nuclear Fusion}\ }\textbf {\bibinfo {volume} {39}},\ \bibinfo {pages} {587}
  (\bibinfo {year} {1999})}\BibitemShut {NoStop}%
\bibitem [{\citenamefont {Kolesnichenko}\ \emph {et~al.}(1992)\citenamefont
  {Kolesnichenko}, \citenamefont {Yakovenko}, \citenamefont {Anderson},
  \citenamefont {Lisak},\ and\ \citenamefont
  {Wising}}]{kolesnichenko1992sawtooth}%
  \BibitemOpen
  \bibfield  {author} {\bibinfo {author} {\bibfnamefont {Y.~I.}\ \bibnamefont
  {Kolesnichenko}}, \bibinfo {author} {\bibfnamefont {Y.~V.}\ \bibnamefont
  {Yakovenko}}, \bibinfo {author} {\bibfnamefont {D.}~\bibnamefont {Anderson}},
  \bibinfo {author} {\bibfnamefont {M.}~\bibnamefont {Lisak}}, \ and\ \bibinfo
  {author} {\bibfnamefont {F.}~\bibnamefont {Wising}},\ }\href@noop {}
  {\bibfield  {journal} {\bibinfo  {journal} {Physical review letters}\
  }\textbf {\bibinfo {volume} {68}},\ \bibinfo {pages} {3881} (\bibinfo {year}
  {1992})}\BibitemShut {NoStop}%
\bibitem [{\citenamefont {Boozer}(1985)}]{boozer1985magnetic}%
  \BibitemOpen
  \bibfield  {author} {\bibinfo {author} {\bibfnamefont {A.~H.}\ \bibnamefont
  {Boozer}},\ }\href@noop {} {\emph {\bibinfo {title} {Magnetic field line
  Hamiltonian}}},\ \bibinfo {type} {Tech. Rep.}\ (\bibinfo  {institution}
  {Princeton Univ.},\ \bibinfo {year} {1985})\BibitemShut {NoStop}%
\bibitem [{\citenamefont {Brouwer}(1911)}]{brouwer1911abbildung}%
  \BibitemOpen
  \bibfield  {author} {\bibinfo {author} {\bibfnamefont {L.~E.~J.}\
  \bibnamefont {Brouwer}},\ }\href@noop {} {\bibfield  {journal} {\bibinfo
  {journal} {Mathematische Annalen}\ }\textbf {\bibinfo {volume} {71}},\
  \bibinfo {pages} {97} (\bibinfo {year} {1911})}\BibitemShut {NoStop}%
\bibitem [{\citenamefont {Greene}(1968)}]{greene1968two}%
  \BibitemOpen
  \bibfield  {author} {\bibinfo {author} {\bibfnamefont {J.~M.}\ \bibnamefont
  {Greene}},\ }\href@noop {} {\bibfield  {journal} {\bibinfo  {journal}
  {Journal of Mathematical Physics}\ }\textbf {\bibinfo {volume} {9}},\
  \bibinfo {pages} {760} (\bibinfo {year} {1968})}\BibitemShut {NoStop}%
\bibitem [{\citenamefont {Greene}(1979)}]{greene1979method}%
  \BibitemOpen
  \bibfield  {author} {\bibinfo {author} {\bibfnamefont {J.~M.}\ \bibnamefont
  {Greene}},\ }\href@noop {} {\bibfield  {journal} {\bibinfo  {journal}
  {Journal of Mathematical Physics}\ }\textbf {\bibinfo {volume} {20}},\
  \bibinfo {pages} {1183} (\bibinfo {year} {1979})}\BibitemShut {NoStop}%
\bibitem [{\citenamefont {Smiet}\ \emph {et~al.}(2019)\citenamefont {Smiet},
  \citenamefont {Kramer},\ and\ \citenamefont {Hudson}}]{smiet2019mapping}%
  \BibitemOpen
  \bibfield  {author} {\bibinfo {author} {\bibfnamefont {C.~B.}\ \bibnamefont
  {Smiet}}, \bibinfo {author} {\bibfnamefont {G.~J.}\ \bibnamefont {Kramer}}, \
  and\ \bibinfo {author} {\bibfnamefont {S.~R.}\ \bibnamefont {Hudson}},\
  }\href@noop {} {\bibfield  {journal} {\bibinfo  {journal} {Plasma Physics and
  Controlled Fusion}\ ,\ \bibinfo {pages} {under review}} (\bibinfo {year}
  {2019})}\BibitemShut {NoStop}%
\bibitem [{\citenamefont {Cheng}(1992)}]{cheng1992kinetic}%
  \BibitemOpen
  \bibfield  {author} {\bibinfo {author} {\bibfnamefont {C.}~\bibnamefont
  {Cheng}},\ }\href@noop {} {\bibfield  {journal} {\bibinfo  {journal} {Physics
  reports}\ }\textbf {\bibinfo {volume} {211}},\ \bibinfo {pages} {1} (\bibinfo
  {year} {1992})}\BibitemShut {NoStop}%
\bibitem [{\citenamefont {Kramer}\ \emph {et~al.}(2013)\citenamefont {Kramer},
  \citenamefont {Budny}, \citenamefont {Bortolon}, \citenamefont {Fredrickson},
  \citenamefont {Fu}, \citenamefont {Heidbrink}, \citenamefont {Nazikian},
  \citenamefont {Valeo},\ and\ \citenamefont
  {Van~Zeeland}}]{kramer2013description}%
  \BibitemOpen
  \bibfield  {author} {\bibinfo {author} {\bibfnamefont {G.~J.}\ \bibnamefont
  {Kramer}}, \bibinfo {author} {\bibfnamefont {R.~V.}\ \bibnamefont {Budny}},
  \bibinfo {author} {\bibfnamefont {A.}~\bibnamefont {Bortolon}}, \bibinfo
  {author} {\bibfnamefont {E.~D.}\ \bibnamefont {Fredrickson}}, \bibinfo
  {author} {\bibfnamefont {G.~Y.}\ \bibnamefont {Fu}}, \bibinfo {author}
  {\bibfnamefont {W.~W.}\ \bibnamefont {Heidbrink}}, \bibinfo {author}
  {\bibfnamefont {R.}~\bibnamefont {Nazikian}}, \bibinfo {author}
  {\bibfnamefont {E.}~\bibnamefont {Valeo}}, \ and\ \bibinfo {author}
  {\bibfnamefont {M.}~\bibnamefont {Van~Zeeland}},\ }\href@noop {} {\bibfield
  {journal} {\bibinfo  {journal} {Plasma Physics and Controlled Fusion}\
  }\textbf {\bibinfo {volume} {55}},\ \bibinfo {pages} {025013} (\bibinfo
  {year} {2013})}\BibitemShut {NoStop}%
\bibitem [{\citenamefont {Nagayama}\ \emph {et~al.}(1996)\citenamefont
  {Nagayama}, \citenamefont {Yamada}, \citenamefont {Park}, \citenamefont
  {Fredrickson}, \citenamefont {Janos}, \citenamefont {McGuire},\ and\
  \citenamefont {Taylor}}]{nagayama1996tomography}%
  \BibitemOpen
  \bibfield  {author} {\bibinfo {author} {\bibfnamefont {Y.}~\bibnamefont
  {Nagayama}}, \bibinfo {author} {\bibfnamefont {M.}~\bibnamefont {Yamada}},
  \bibinfo {author} {\bibfnamefont {W.}~\bibnamefont {Park}}, \bibinfo {author}
  {\bibfnamefont {E.}~\bibnamefont {Fredrickson}}, \bibinfo {author}
  {\bibfnamefont {A.}~\bibnamefont {Janos}}, \bibinfo {author} {\bibfnamefont
  {K.}~\bibnamefont {McGuire}}, \ and\ \bibinfo {author} {\bibfnamefont
  {G.}~\bibnamefont {Taylor}},\ }\href@noop {} {\bibfield  {journal} {\bibinfo
  {journal} {Physics of Plasmas}\ }\textbf {\bibinfo {volume} {3}},\ \bibinfo
  {pages} {1647} (\bibinfo {year} {1996})}\BibitemShut {NoStop}%
\bibitem [{\citenamefont {Boozer}(2014)}]{boozer2014formation}%
  \BibitemOpen
  \bibfield  {author} {\bibinfo {author} {\bibfnamefont {A.~H.}\ \bibnamefont
  {Boozer}},\ }\href@noop {} {\bibfield  {journal} {\bibinfo  {journal}
  {Physics of Plasmas}\ }\textbf {\bibinfo {volume} {21}},\ \bibinfo {pages}
  {072907} (\bibinfo {year} {2014})}\BibitemShut {NoStop}%
\bibitem [{\citenamefont {Huang}\ \emph {et~al.}(2014)\citenamefont {Huang},
  \citenamefont {Bhattacharjee},\ and\ \citenamefont
  {Boozer}}]{huang2014rapid}%
  \BibitemOpen
  \bibfield  {author} {\bibinfo {author} {\bibfnamefont {Y.-M.}\ \bibnamefont
  {Huang}}, \bibinfo {author} {\bibfnamefont {A.}~\bibnamefont
  {Bhattacharjee}}, \ and\ \bibinfo {author} {\bibfnamefont {A.~H.}\
  \bibnamefont {Boozer}},\ }\href@noop {} {\bibfield  {journal} {\bibinfo
  {journal} {The Astrophysical Journal}\ }\textbf {\bibinfo {volume} {793}},\
  \bibinfo {pages} {106} (\bibinfo {year} {2014})}\BibitemShut {NoStop}%
\bibitem [{\citenamefont {Chapman}\ \emph {et~al.}(2007)\citenamefont
  {Chapman}, \citenamefont {Pinches}, \citenamefont {Graves}, \citenamefont
  {Akers}, \citenamefont {Appel}, \citenamefont {Budny}, \citenamefont {Coda},
  \citenamefont {Conway}, \citenamefont {De~Bock}, \citenamefont {Eriksson}
  \emph {et~al.}}]{chapman2007physics}%
  \BibitemOpen
  \bibfield  {author} {\bibinfo {author} {\bibfnamefont {I.~T.}\ \bibnamefont
  {Chapman}}, \bibinfo {author} {\bibfnamefont {S.~D.}\ \bibnamefont
  {Pinches}}, \bibinfo {author} {\bibfnamefont {J.~P.}\ \bibnamefont {Graves}},
  \bibinfo {author} {\bibfnamefont {R.~J.}\ \bibnamefont {Akers}}, \bibinfo
  {author} {\bibfnamefont {L.~C.}\ \bibnamefont {Appel}}, \bibinfo {author}
  {\bibfnamefont {R.~V.}\ \bibnamefont {Budny}}, \bibinfo {author}
  {\bibfnamefont {S.}~\bibnamefont {Coda}}, \bibinfo {author} {\bibfnamefont
  {N.~J.}\ \bibnamefont {Conway}}, \bibinfo {author} {\bibfnamefont
  {M.}~\bibnamefont {De~Bock}}, \bibinfo {author} {\bibfnamefont {L.~G.}\
  \bibnamefont {Eriksson}},  \emph {et~al.},\ }\href@noop {} {\bibfield
  {journal} {\bibinfo  {journal} {Plasma Physics and Controlled Fusion}\
  }\textbf {\bibinfo {volume} {49}},\ \bibinfo {pages} {B385} (\bibinfo {year}
  {2007})}\BibitemShut {NoStop}%
\bibitem [{\citenamefont {Chapman}\ \emph {et~al.}(2006)\citenamefont
  {Chapman}, \citenamefont {Hender}, \citenamefont {Saarelma}, \citenamefont
  {Sharapov}, \citenamefont {Akers}, \citenamefont {Conway}, \citenamefont
  {Team} \emph {et~al.}}]{chapman2006effect}%
  \BibitemOpen
  \bibfield  {author} {\bibinfo {author} {\bibfnamefont {I.~T.}\ \bibnamefont
  {Chapman}}, \bibinfo {author} {\bibfnamefont {T.~C.}\ \bibnamefont {Hender}},
  \bibinfo {author} {\bibfnamefont {S.}~\bibnamefont {Saarelma}}, \bibinfo
  {author} {\bibfnamefont {S.~E.}\ \bibnamefont {Sharapov}}, \bibinfo {author}
  {\bibfnamefont {R.~J.}\ \bibnamefont {Akers}}, \bibinfo {author}
  {\bibfnamefont {N.~J.}\ \bibnamefont {Conway}}, \bibinfo {author}
  {\bibfnamefont {M.}~\bibnamefont {Team}},  \emph {et~al.},\ }\href@noop {}
  {\bibfield  {journal} {\bibinfo  {journal} {Nuclear fusion}\ }\textbf
  {\bibinfo {volume} {46}},\ \bibinfo {pages} {1009} (\bibinfo {year}
  {2006})}\BibitemShut {NoStop}%
\bibitem [{\citenamefont {Cole}\ \emph {et~al.}(2014)\citenamefont {Cole},
  \citenamefont {Mishchenko}, \citenamefont {K{\"o}nies}, \citenamefont
  {Kleiber},\ and\ \citenamefont {Borchardt}}]{cole2014fluid}%
  \BibitemOpen
  \bibfield  {author} {\bibinfo {author} {\bibfnamefont {M.}~\bibnamefont
  {Cole}}, \bibinfo {author} {\bibfnamefont {A.}~\bibnamefont {Mishchenko}},
  \bibinfo {author} {\bibfnamefont {A.}~\bibnamefont {K{\"o}nies}}, \bibinfo
  {author} {\bibfnamefont {R.}~\bibnamefont {Kleiber}}, \ and\ \bibinfo
  {author} {\bibfnamefont {M.}~\bibnamefont {Borchardt}},\ }\href@noop {}
  {\bibfield  {journal} {\bibinfo  {journal} {Physics of Plasmas}\ }\textbf
  {\bibinfo {volume} {21}},\ \bibinfo {pages} {072123} (\bibinfo {year}
  {2014})}\BibitemShut {NoStop}%
\bibitem [{\citenamefont {Porcelli}(1991)}]{porcelli1991fast}%
  \BibitemOpen
  \bibfield  {author} {\bibinfo {author} {\bibfnamefont {F.}~\bibnamefont
  {Porcelli}},\ }\href@noop {} {\bibfield  {journal} {\bibinfo  {journal}
  {Plasma Physics and Controlled Fusion}\ }\textbf {\bibinfo {volume} {33}},\
  \bibinfo {pages} {1601} (\bibinfo {year} {1991})}\BibitemShut {NoStop}%
\bibitem [{\citenamefont {Connor}\ \emph {et~al.}(2012)\citenamefont {Connor},
  \citenamefont {Hastie},\ and\ \citenamefont {Zocco}}]{connor2012unified}%
  \BibitemOpen
  \bibfield  {author} {\bibinfo {author} {\bibfnamefont {J.~W.}\ \bibnamefont
  {Connor}}, \bibinfo {author} {\bibfnamefont {R.~J.}\ \bibnamefont {Hastie}},
  \ and\ \bibinfo {author} {\bibfnamefont {A.}~\bibnamefont {Zocco}},\
  }\href@noop {} {\bibfield  {journal} {\bibinfo  {journal} {Plasma Physics and
  Controlled Fusion}\ }\textbf {\bibinfo {volume} {54}},\ \bibinfo {pages}
  {035003} (\bibinfo {year} {2012})}\BibitemShut {NoStop}%
\bibitem [{\citenamefont {Angioni}\ \emph {et~al.}(2002)\citenamefont
  {Angioni}, \citenamefont {Pochelon}, \citenamefont {Gorelenkov},
  \citenamefont {McClements}, \citenamefont {Sauter}, \citenamefont {Budny},
  \citenamefont {De~Vries}, \citenamefont {Howell}, \citenamefont {Mantsinen},
  \citenamefont {Nave} \emph {et~al.}}]{angioni2002neutral}%
  \BibitemOpen
  \bibfield  {author} {\bibinfo {author} {\bibfnamefont {C.}~\bibnamefont
  {Angioni}}, \bibinfo {author} {\bibfnamefont {A.}~\bibnamefont {Pochelon}},
  \bibinfo {author} {\bibfnamefont {N.~N.}\ \bibnamefont {Gorelenkov}},
  \bibinfo {author} {\bibfnamefont {K.~G.}\ \bibnamefont {McClements}},
  \bibinfo {author} {\bibfnamefont {O.}~\bibnamefont {Sauter}}, \bibinfo
  {author} {\bibfnamefont {R.~V.}\ \bibnamefont {Budny}}, \bibinfo {author}
  {\bibfnamefont {P.~C.}\ \bibnamefont {De~Vries}}, \bibinfo {author}
  {\bibfnamefont {D.~F.}\ \bibnamefont {Howell}}, \bibinfo {author}
  {\bibfnamefont {M.}~\bibnamefont {Mantsinen}}, \bibinfo {author}
  {\bibfnamefont {M.~F.~F.}\ \bibnamefont {Nave}},  \emph {et~al.},\
  }\href@noop {} {\bibfield  {journal} {\bibinfo  {journal} {Plasma physics and
  controlled fusion}\ }\textbf {\bibinfo {volume} {44}},\ \bibinfo {pages}
  {205} (\bibinfo {year} {2002})}\BibitemShut {NoStop}%
\bibitem [{\citenamefont {Campbell}\ \emph {et~al.}(1988)\citenamefont
  {Campbell}, \citenamefont {Start}, \citenamefont {Wesson}, \citenamefont
  {Bartlett}, \citenamefont {Bhatnagar}, \citenamefont {Bures}, \citenamefont
  {Cordey}, \citenamefont {Cottrell}, \citenamefont {Dupperex}, \citenamefont
  {Edwards} \emph {et~al.}}]{campbell1988stabilization}%
  \BibitemOpen
  \bibfield  {author} {\bibinfo {author} {\bibfnamefont {D.~J.}\ \bibnamefont
  {Campbell}}, \bibinfo {author} {\bibfnamefont {D.~F.~H.}\ \bibnamefont
  {Start}}, \bibinfo {author} {\bibfnamefont {J.~A.}\ \bibnamefont {Wesson}},
  \bibinfo {author} {\bibfnamefont {D.~V.}\ \bibnamefont {Bartlett}}, \bibinfo
  {author} {\bibfnamefont {V.~P.}\ \bibnamefont {Bhatnagar}}, \bibinfo {author}
  {\bibfnamefont {M.}~\bibnamefont {Bures}}, \bibinfo {author} {\bibfnamefont
  {J.~G.}\ \bibnamefont {Cordey}}, \bibinfo {author} {\bibfnamefont {G.~A.}\
  \bibnamefont {Cottrell}}, \bibinfo {author} {\bibfnamefont {P.~A.}\
  \bibnamefont {Dupperex}}, \bibinfo {author} {\bibfnamefont {A.~W.}\
  \bibnamefont {Edwards}},  \emph {et~al.},\ }\href@noop {} {\bibfield
  {journal} {\bibinfo  {journal} {Physical review letters}\ }\textbf {\bibinfo
  {volume} {60}},\ \bibinfo {pages} {2148} (\bibinfo {year}
  {1988})}\BibitemShut {NoStop}%
\bibitem [{\citenamefont {Bernabei}\ \emph {et~al.}(2001)\citenamefont
  {Bernabei}, \citenamefont {Budny}, \citenamefont {Fredrickson}, \citenamefont
  {Gorelenkov}, \citenamefont {Hosea}, \citenamefont {Phillips}, \citenamefont
  {White}, \citenamefont {Wilson}, \citenamefont {Petty}, \citenamefont
  {Pinsker} \emph {et~al.}}]{bernabei2001combined}%
  \BibitemOpen
  \bibfield  {author} {\bibinfo {author} {\bibfnamefont {S.}~\bibnamefont
  {Bernabei}}, \bibinfo {author} {\bibfnamefont {R.~V.}\ \bibnamefont {Budny}},
  \bibinfo {author} {\bibfnamefont {E.~D.}\ \bibnamefont {Fredrickson}},
  \bibinfo {author} {\bibfnamefont {N.~N.}\ \bibnamefont {Gorelenkov}},
  \bibinfo {author} {\bibfnamefont {J.~C.}\ \bibnamefont {Hosea}}, \bibinfo
  {author} {\bibfnamefont {C.~K.}\ \bibnamefont {Phillips}}, \bibinfo {author}
  {\bibfnamefont {R.~B.}\ \bibnamefont {White}}, \bibinfo {author}
  {\bibfnamefont {J.~R.}\ \bibnamefont {Wilson}}, \bibinfo {author}
  {\bibfnamefont {C.~C.}\ \bibnamefont {Petty}}, \bibinfo {author}
  {\bibfnamefont {R.~I.}\ \bibnamefont {Pinsker}},  \emph {et~al.},\
  }\href@noop {} {\bibfield  {journal} {\bibinfo  {journal} {Nuclear Fusion}\
  }\textbf {\bibinfo {volume} {41}},\ \bibinfo {pages} {513} (\bibinfo {year}
  {2001})}\BibitemShut {NoStop}%
\bibitem [{\citenamefont {Chang}\ \emph {et~al.}(1996)\citenamefont {Chang},
  \citenamefont {Park}, \citenamefont {Fredrickson}, \citenamefont {Batha},
  \citenamefont {Bell}, \citenamefont {Bell}, \citenamefont {Budny},
  \citenamefont {Bush}, \citenamefont {Janos}, \citenamefont {Levinton} \emph
  {et~al.}}]{chang1996off}%
  \BibitemOpen
  \bibfield  {author} {\bibinfo {author} {\bibfnamefont {Z.}~\bibnamefont
  {Chang}}, \bibinfo {author} {\bibfnamefont {W.}~\bibnamefont {Park}},
  \bibinfo {author} {\bibfnamefont {E.~D.}\ \bibnamefont {Fredrickson}},
  \bibinfo {author} {\bibfnamefont {S.~H.}\ \bibnamefont {Batha}}, \bibinfo
  {author} {\bibfnamefont {M.~G.}\ \bibnamefont {Bell}}, \bibinfo {author}
  {\bibfnamefont {R.}~\bibnamefont {Bell}}, \bibinfo {author} {\bibfnamefont
  {R.~V.}\ \bibnamefont {Budny}}, \bibinfo {author} {\bibfnamefont {C.~E.}\
  \bibnamefont {Bush}}, \bibinfo {author} {\bibfnamefont {A.}~\bibnamefont
  {Janos}}, \bibinfo {author} {\bibfnamefont {F.~M.}\ \bibnamefont {Levinton}},
   \emph {et~al.},\ }\href@noop {} {\bibfield  {journal} {\bibinfo  {journal}
  {Physical review letters}\ }\textbf {\bibinfo {volume} {77}},\ \bibinfo
  {pages} {3553} (\bibinfo {year} {1996})}\BibitemShut {NoStop}%
\end{thebibliography}
\end{document}